\documentclass[nofootinbib,prd,aps,superscriptaddress,preprintnumbers,preprint]{revtex4}
\usepackage{epsfig,amsmath,amssymb,euscript}
\usepackage{color}
\usepackage{accents}
\usepackage{hyperref}
\def\beq{\begin{equation}}
\def\eeq{\end{equation}}

\begin{document}

\title{The ultimate theoretical error on $\gamma$ from $B\to D K$ decays}

\def\Cincy{Department of Physics, University of Cincinnati, Cincinnati, Ohio 45221,USA}
\def\NIPNE{National Institute for Physics and Nuclear Engineering, 
Department of Particle Physics, 077125 Bucharest, Romania}
\author{Joachim Brod}
\email[Electronic address:]{brodjm@ucmail.uc.edu}
\affiliation{\Cincy}
\author{Jure Zupan}
\email[Electronic address:]{zupanje@ucmail.uc.edu}
\affiliation{\Cincy}

\vspace{1.0cm}
\begin{abstract}
\vspace{0.2cm}\noindent The angle $\gamma$ of the standard CKM
unitarity triangle can be determined from $B\to D K$ decays with a
very small irreducible theoretical error, which is only due to
second-order electroweak corrections. We study these contributions and
estimate that their impact on the $\gamma$ determination is to
introduce a shift $|\delta \gamma | \lesssim {\mathcal O}(10^{-7})$,
well below any present or planned future experiment.
\end{abstract}
\maketitle

\section{Introduction}

The determination of the standard CKM unitarity triangle angle $\gamma
\equiv \arg(-V_{ud}V_{ub}^*/V_{cd}V_{cb}^*)$ from $B\to DK$ and $B\to
\bar{D} K$ decays is theoretically extremely clean. The reason is that
the $B\to D K$ transitions receive contributions only from tree
operators, and none from penguin operators. Furthermore, all the
relevant matrix elements can be obtained from data if enough $D$-decay
channels are measured.  The sensitivity to $\gamma$ comes from the
interference of $b\to c\bar u s$ and $b\to u\bar c s$ decay
amplitudes, which have a relative weak phase $\gamma$,
cf. Fig. \ref{fig:tree}.  These quark-level transitions mediate $B^-
\to D^0 K^-$ and $B^- \to \bar D^0 K^-$ decays, respectively. The
$D^0$ and $\bar D^0$ subsequently decay into a common final state $f$,
which allows the two decay channels to interfere. Several variants of
the method have been proposed, distinguished by the final state $f$:
i) $f$ can be a CP eigenstate such as $K_S\pi^0$ and $K_S\phi$
\cite{Gronau:1990ra,Gronau:1991dp}, ii) a flavor state such as
$K^+\pi^-$ and $K^{*+} \rho^-$ \cite{Atwood:1996ci}, or iii) a
multibody state such as $K_S\pi^+\pi^-, \pi^+\pi^-\pi^0$
\cite{Giri:2003ty,Grossman:2002aq,Bondar:2005ki}. Other possibilities
include the decays of neutral $B$ mesons, $B^0$ and $B_s$,
\cite{Kayser:1999bu,Gronau:2004gt}, multibody $B$ decays
\cite{Aleksan:2002mh,Gershon:2009qr,Gershon:2008pe,Gershon:2009qc} and
$D^*$ or $D^{**}$ decays \cite{Bondar:2004bi,Sinha:2004ct} (see also
the reviews in \cite{Zupan:2007zz} and the current combination of LHCb
measurements in \cite{Aaij:2013zfa}).

The above set of methods has several sources of theoretical
errors. Most of them can be reduced once more statistics becomes
available. For instance, in the past the $D\to K_S\pi^+\pi^-$ Dalitz
plot needed to be modeled using a sum of Breit-Wigner resonances or
using the K-matrix formalism. Utilizing the data from entangled
$\psi(3770\to D\bar D)$ decays measured at CLEO-c \cite{Libby:2010nu}
and BES-III, this uncertainty can in principle be completely avoided
\cite{Grossman:2002aq}. The related error is now statistics-dominated
\cite{Aihara:2012aw,Aaij:2012hu}.

Other sources of reducible uncertainties are $D-\bar D$ mixing and
$K-\bar K$ mixing (for final states with $K_S$). Both of these effects
can be included trivially by modifying the expressions for the decay
amplitudes, taking meson mixing into account, and then using
experimentally measured mixing parameters \cite{Silva:1999bd}. The
effect of $D-\bar D$ mixing is most significant if the $D$ decay
information comes from entangled $\psi(3770\to D\bar D)$ decays. The
shift in $\gamma$ is then linear in $x_D, y_D$, giving $\Delta \gamma
\leqslant 2.9^\circ$ \cite{Bondar:2010qs} (see also
\cite{Rama:2013voa}). For flavor-tagged $D$ decays (i.e. from $D^*\to
D\pi)$ the effect is quadratic in $x_D, y_D$ and thus much smaller
\cite{Grossman:2005rp}. Similarly, for $\gamma$ extraction from
untagged $B_s\to D \phi$ decays the inclusion of $\Delta \Gamma_s$ can
be important and can be achieved once $\Delta \Gamma_s$ is well
measured \cite{Gronau:2007bh}. In the extraction of $\gamma$ from
$B\to DK$, CP violation in the $D$ system was usually neglected.  Even
if this assumption is relaxed, it is still possible to extract
$\gamma$ by appropriately modifying the expressions for the decay
amplitudes (and using the fact that in the Cabibbo allowed $D$ decays
there is no direct CP violation) \cite{Aaij:2013zfa, LHCb-measur,
  Wang:2012ie, Martone:2012nj, Bhattacharya:2013vc, Bondar:2013jxa}.

Yet another source of reducible theory error are QED radiative
corrections to the decay widths. The uncertainties from this source
are expected to be below present experimental sensitivity on $\gamma$
so that not much work has been done on them. Since the corrections are
CP conserving they can be reabsorbed in the CP-even measured hadronic
quantities and would not affect $\gamma$, as long as in the
measurements the radiative corrections are treated consistently
between different decay modes.

The first irreducible theory error on $\gamma$ thus comes from
higher-order electroweak corrections.  This error cannot be eliminated
using just experimental information and may well represent the
ultimate precision of the $\gamma$ determination from $B\to DK$
decays. The resulting uncertainty was estimated using scaling arguments
in Ref.~\cite{Zupan:2011mn} and found to be of the order of $\delta
\gamma/\gamma\sim {\mathcal O}(10^{-6})$.  In this paper we perform a
more careful analysis, and find that the induced uncertainty is in
fact most probably even an order of magnitude smaller. The one-loop
electroweak corrections give rise to local and nonlocal
contributions. We estimate the size of the local contributions using
naive factorization and obtain $\delta \gamma/\gamma\lesssim {\mathcal
  O}(10^{-7})$. The nonlocal contributions are more difficult to
estimate, but naively one expects that they are not significantly
larger than the local ones.

The paper is organized as follows. In Section \ref{sec:2} we give a
brief discussion of electroweak corrections for $B\to DK$ decays with
a focus on the $\gamma$ extraction. We also give numerical estimates
for the shift, $\delta\gamma$, utilizing the analytic results of
Section \ref{sec:3}, where further details of the calculation are
given. Finally, we conclude in Section \ref{Conclusions}.

\section{The shift in $\gamma$ from $B\to DK$ due to electroweak corrections}
\label{sec:2}

\begin{figure}
    \centering
   \includegraphics[width=4cm]{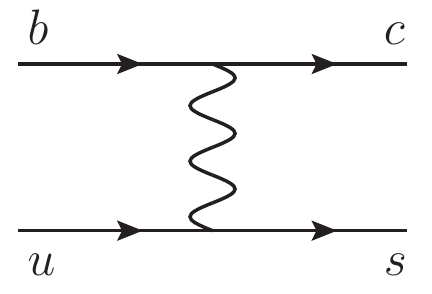}  
~~~~~~~~~~~~~~~~~~~\includegraphics[width=4cm]{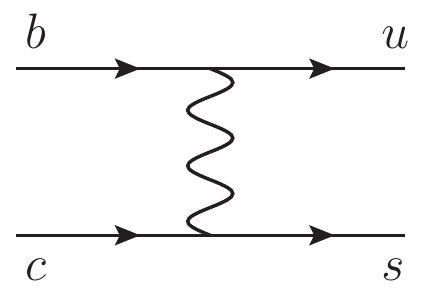}  \\
    \caption{
Tree contributions (with single $W$ exchange) that mediate $b\to c
\bar u s$ (left) and $b\to u\bar c s$ (right) quark-level processes,
which lead to $B^-\to D^0 K^-$ and $B^-\to \bar D^0 K^-$ decays,
respectively. } 
\label{fig:tree}
\end{figure}

The measurement of $\gamma$ in $B\to DK$ decays is based on the
interference between the tree-level $b \to c \bar u s$ and $b\to u
\bar c s$ mediated processes, cf. Fig. \ref{fig:tree}. The
sensitivity to the weak phase $\gamma$ enters through the amplitude
ratio
\begin{eqnarray}\label{rB}
r_B e^{i(\delta_B-\gamma)} = \frac{A(B^- \to \bar D^0 K^-)}{A(B^- \to D^0 K^-)},
\end{eqnarray}
where $\delta_B=(114.8\pm9.4 )^\circ$ is a strong phase, and
$r_B=0.0956\pm0.0063$ reflects the CKM and color suppression of the
amplitude $A(B^-\to \bar D^0 K^-)$ relative to the amplitude $A(B^-
\to D^0 K^-)$ \cite{Charles:2011va}. Here and below we focus on the
charged $B^-\to D K^-$ and $B^-\to \bar DK^-$ decays. The results can
be readily adapted also to other $B\to DK$ or $B_s\to D_s K$ decays
used for extraction of $\gamma$.

The expression \eqref{rB} is valid only at leading order in the weak
interactions, ${\mathcal O}(G_F)$, when
both the $b \to c \bar u s$ and $b\to u \bar c s$ transitions are
mediated by the tree-level processes.  At this order the two processes
are described by the usual nonleptonic weak effective Hamiltonians
\begin{eqnarray}\label{1}
{\cal H}_{\bar c u}^{(0)} = 
\frac{G_F}{\sqrt2} V_{cb} V^*_{us}\big[ C_1(\mu) Q_1^{\bar c u}
+ C_2(\mu) Q_2^{\bar c u}  \big], \\
\label{2}
{\cal H}_{\bar u c}^{(0)} = 
\frac{G_F}{\sqrt2} V_{ub} V^*_{cs}\big[ C_1(\mu) Q_1^{\bar u c}
+ C_2(\mu) Q_2^{\bar u c}  \big], 
\end{eqnarray}
where the four-fermion operators are
\begin{align}
Q_1^{\bar c u}&= (\bar cb)_{V-A}(\bar su)_{V-A}, \qquad
Q_2^{\bar c u}= (\bar s b)_{V-A}(\bar c u)_{V-A}  , \label{Q12}\\
 Q_1^{\bar u c}&= (\bar ub)_{V-A}(\bar sc)_{V-A},\qquad
Q_2^{\bar u c}=(\bar s b)_{V-A}(\bar u c)_{V-A} . 
\end{align}
Above we have used the short-hand notation $(\bar cb)_{V-A}(\bar
su)_{V-A}\equiv\big(\bar c\gamma^\mu(1-\gamma_5)b\big)\,\big(\bar
s\gamma_\mu(1-\gamma_5)u\big)$, and similarly for the other quark
flavors. The scale at which the Wilson coefficients are evaluated is
close to the $b$ quark mass, $\mu\sim m_b$, with $C_1(m_b)=1.10$, and
$C_2(m_b)=-0.24$ at leading-log order~\cite{Buchalla:1995vs}, for
$m_b(m_b) = 4.163\,\text{GeV}$~\cite{Chetyrkin:2009fv} and
$\alpha_S(M_Z)=0.1184$~\cite{Beringer:1900zz}.  The decay amplitudes
in Eq. \eqref{rB} are then given at leading order in the electroweak
expansion by
\begin{equation}\label{eqA}
A(B^- \to \bar D^0 K^-)=\langle \bar D^0 K^-| {\cal H}_{\bar u
  c}^{(0)}| B^-\rangle, \quad\text{and}\quad A(B^- \to D^0
K^-)=\langle D^0 K^-| {\cal H}_{\bar c u}^{(0)}| B^-\rangle. 
\end{equation}

\begin{figure}
    \centering
\includegraphics[height=3cm]{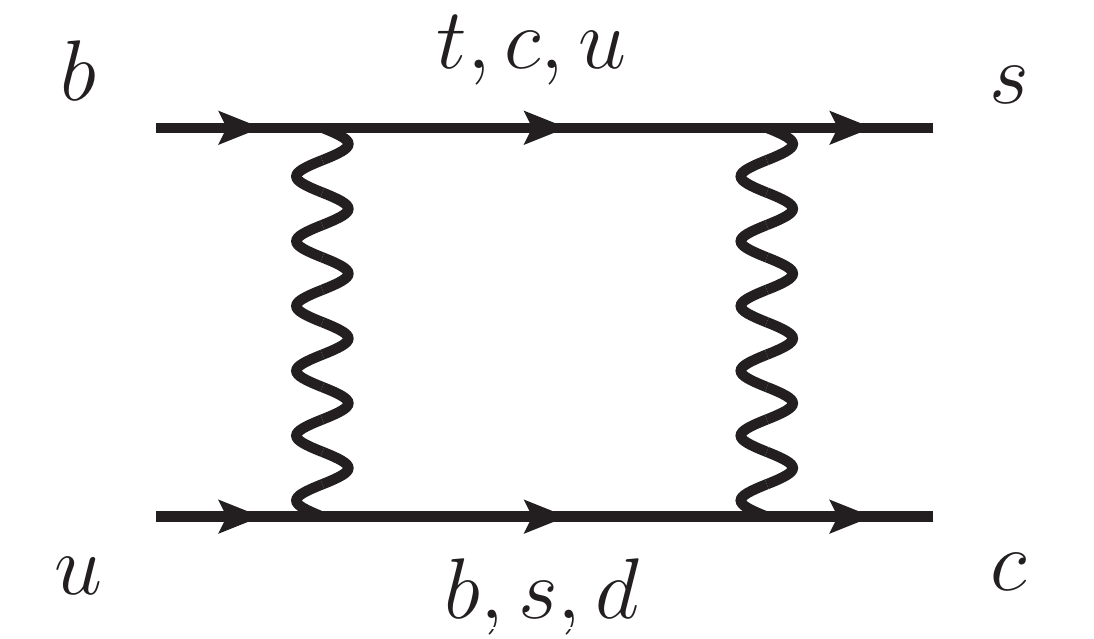} ~~~ 
\includegraphics[height=2.5cm]{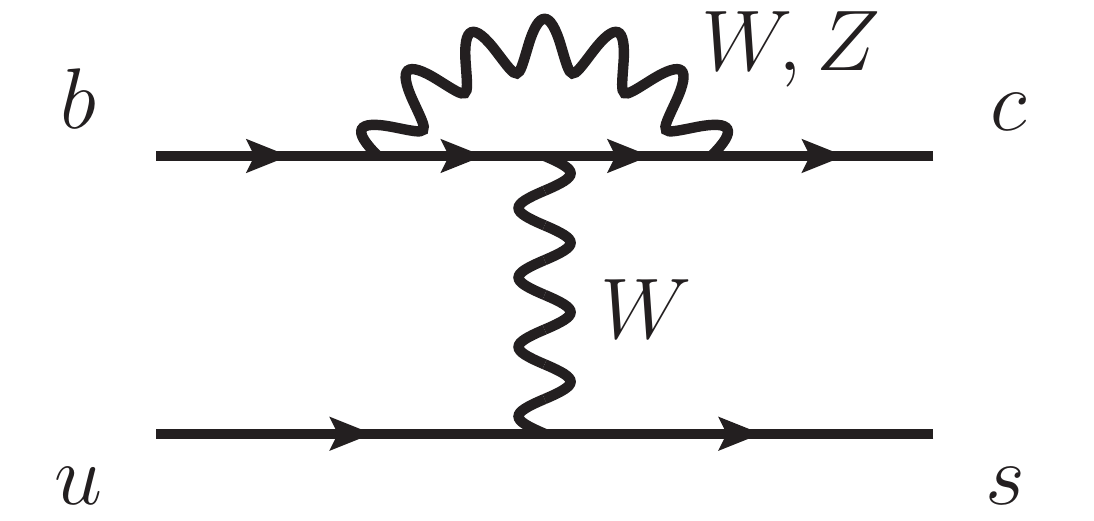}  
\caption{ The electroweak corrections to $b\to c \bar u s$ process at
  order ${\mathcal O}(g^4)$, the box diagram (left) 
  and vertex correction (right). Similar diagrams appear in $b\to u
  \bar c s$ processes. }
\label{fig:ewcorr}
\end{figure}

At second order in the weak interactions, ${\mathcal O}(G_F^2)$, there
are corrections to \eqref{rB} and \eqref{eqA} from $W$ box diagrams,
and from vertex corrections, shown in Fig. \ref{fig:ewcorr}, and from
double penguin diagrams.  In addition there are also self-energy
diagrams for the $W$-propagator and wave function renormalization
diagrams for external legs, which however have exactly the same CKM
structure as the leading order contributions and thus do not affect
the $\gamma$ extraction.  The same is true of the vertex corrections
due to a $Z$ or $W$ loop, shown in Fig.~\ref{fig:ewcorr} (right),
which correct the CKM matrix at one-loop.  The double penguin
insertions are two-loop and are thus subleading, as can be easily
checked from the small sizes of the respective Wilson
coefficients. They are safely neglected in the following.

The leading effect on extracted $ \gamma$ at ${\mathcal O}(G_F^2)$
then comes from the box diagram in Fig.  \ref{fig:ewcorr} (left). The
dominant contribution is effectively due to the top and bottom quark
running in the loop, as we show in the next section. The CKM structure
of the box diagram is different from that of the ${\mathcal O}(G_F)$
tree contribution and is given, for the $b \to c s\bar u$ transition,
by
\begin{eqnarray}
b \to c s\bar u: \mbox{ tree level} \sim V_{cb} V^*_{us}\,,\qquad
\mbox{ box diagram } \sim (V_{tb} V^*_{ts})(V_{cb} V^*_{ub})\,. \label{btocsbaru}
\end{eqnarray}
Since the weak phases of the two contributions are different, this
results in a shift $\delta \gamma$ in the extracted value of
$\gamma$.

A similar higher-order electroweak diagram contributes also to the
$b\to u \bar c s$ transition, which is given by exchanging the
external $u$ and $c$ quarks in Fig.  \ref{fig:ewcorr} (left). Again,
the dominant contribution is effectively due to the top and bottom
quark running in the loop, so that the CKM factors are
\begin{eqnarray}
b \to u s\bar c: \mbox{ tree level} \sim V_{ub} V^*_{cs}\,,\qquad
\mbox{ box diagram } \sim (V_{tb} V^*_{ts})(V_{ub} V^*_{cb})\,.\label{btousbarc}
\end{eqnarray}
In this case the weak phases of the LO and NLO contributions are the
same to a very good approximation, so that the electroweak
contributions do not induce a shift in $\gamma$.

Keeping only the local part of the box diagram, the relevant change to
the effective weak Hamiltonian is very simple. The structure of the
CKM coefficients in \eqref{btocsbaru} and \eqref{btousbarc} is such
that all the corrections relevant for the $\gamma$ extraction are in
the ${\cal H}_{\bar c u}$ effective weak Hamiltonian Eq.~\eqref{1},
which at ${\mathcal O}(G_F^2)$ takes the form
\begin{equation}\label{H1}
{\cal H}_{\bar c u}^{(1)} = \frac{G_F}{\sqrt2} V_{cb} V^*_{us}\big[
  \big(C_1(\mu)+\Delta C_1(\mu)\big) Q_1^{\bar c u} +
  \big(C_2(\mu)+\Delta C_2(\mu)\big) Q_2^{\bar c u} \big].
\end{equation}
The Wilson coefficients $C_{1,2}(\mu)$ are the same Wilson
coefficients as in Eqs.~(\ref{1}) and (\ref{2}), while $\Delta
C_{1,2}(\mu)$ are calculable corrections. They depend on the CKM
elements and carry a weak phase $\gamma$. They therefore have a
different weak phase than $C_{1,2}(\mu)$, which in our phase
convention are real. This introduces a shift in $\delta\gamma$ in the
extraction of the weak phase $\gamma$ from $B\to DK$ decays. This
shift represent the ultimate theory error on the measurement of
$\gamma$.

Defining the ratio of matrix elements for the two relevant operators
\begin{equation}\label{rA}
r_A\equiv\frac{\langle K^-D^0| Q_2^{\bar cu}|B^-\rangle}{\langle K^-
  D^0|Q_1^{\bar c u}|B^-\rangle},
\end{equation}
the shift in the ratio $r_B$, Eq.~\eqref{rB}, is 
\begin{equation}
r_B e^{i(\delta_B-\gamma)}\to r_B e^{i(\delta_B-\gamma)}
\Big(1 - \frac{\Delta C_1}{C_1 + C_2 r_A} - \frac{\Delta C_2}{C_1/r_A + C_2} \Big),
\end{equation}
where we expanded in the small corrections $\Delta C_1$, $\Delta C_2$
to linear order. The resulting shift in the extracted value of
$\gamma$ is
\begin{equation}
\delta \gamma =  \frac{{\rm Im}(\Delta C_1)}{C_1 + C_2 r_A} +
\frac{{\rm Im}(\Delta C_2)}{C_1/r_A + C_2}\, .
\end{equation}
The size of the corrections $\Delta C_{1,2}$ will be calculated in the
next section, while here we only quote the numerical results. The
unresummed result for ${\rm Im}(\Delta C_2)$,
cf. Eq. \eqref{eq:unresum} below, is
\begin{equation}
{\rm Im}(\Delta C_1) = 0 \, , \quad {\rm Im}(\Delta C_2)=(5.9\pm0.3)
\cdot10^{-8} \times \sin \gamma\, , 
\end{equation}
where the error only reflects the experimental errors due to the input
parameters. The results with $\log(m_b/M_W)$ resummed,
cf. Eq. \eqref{DeltaC12-resum} below, are
\begin{equation}
{\rm Im}(\Delta C_1)=(4.5\pm0.2)\cdot 10^{-9}\times \sin\gamma\, ,
\quad {\rm Im}(\Delta C_2)=(4.3\pm0.2)\cdot 10^{-8}\times \sin\gamma\,. 
\end{equation}
In order to obtain $\delta \gamma$ we also need to estimate the ratio
of the matrix elements, $r_A$, in \eqref{rA}. In naive factorization
this ratio is
\begin{equation}\label{rAnumeric}
r_A=\frac{f_D F_0^{B\to K}(0)}{f_K F_0^{B\to D}(0)}=0.4,
\end{equation}
where we used $f_D=0.214$ GeV \cite{Laiho:2009eu}, $F_0^{B\to
  K}(0)=0.34$ \cite{Beneke:2003zv}, $f_K=0.16$ GeV, $F_0^{B\to
  D}(0)=1.12$ \cite{Neubert:1993mb}. In Eq. \eqref{rAnumeric} we only
quote the central value, since the error on this estimate is bigger
than the errors on the form factors themselves. However, we do not
expect the error on the estimate of $r_A$ in \eqref{rAnumeric} to be
bigger than a factor of a few.

Using this and setting $\gamma=68^\circ$ for definiteness, we obtain
the estimate for the shift $\delta \gamma$, 
\begin{equation}
\delta \gamma\simeq 2.0 \cdot 10^{-8}\, 
\end{equation}
where to this accuracy the resummed expressions for $\Delta
C_{1,2}$ (with nonlocal contributions neglected) and unresummed results coincide.  An uncertainty of at most
an additional factor of a few can be expected on the above estimate,
so that we can conclude that the ultimate theoretical error on
$\gamma$ measurement is safely below
\begin{equation}
|\delta \gamma|\lesssim 10^{-7}\, .
\end{equation}

In the next section we derive the analytic expressions for $\Delta
C_{1,2}(\mu)$, and then draw our conclusions in
Section~\ref{Conclusions}.

\section{Corrections to the electroweak Hamiltonian}
\label{sec:3}

In this section we consider the $b\to c \bar u s$ box diagram,
Fig. \ref{fig:ewcorr} (left), in detail. The results can be readily
adapted to the $b\to u \bar c s$ case by exchanging the external
quarks and adjusting the CKM factors.  The diagram in
Fig. \ref{fig:ewcorr} (left) is superficially similar to the box
diagrams contributing to $\bar K^0-K^0$ and $\bar B_{(s)}^0-B_{(s)}^0$
mixing \cite{Buchalla:1995vs}, and to $b\to ss\bar d, dd\bar s$ decays
\cite{Pirjol:2009vz}. The difference is that the box diagram in
Fig. \ref{fig:ewcorr} (left) has both up- and down-quarks running in
the loop, in contrast to the case of $\bar K^0-K^0$ and $\bar
B_{(s)}^0-B_{(s)}^0$ mixing where both quarks in the loop are of
up-type.

We will calculate the shift $\delta \gamma$ in two ways -- first by
keeping only the $\log(m_b/M_W)$ enhanced local contribution, but
without resumming it. Subsequently we will resum this log. In the
first case we will take $b$, $t$ and $W$ in the loop to be heavy and
integrate them out at $\mu\sim M_W$.
In this way one obtains the local operator part of the effective field
theory (EFT) with only the light quarks, $u,d,s,c$, and an external
non-dynamical $b$-quark field.
Keeping only the local operators in EFT is a crude approximation that
does, however, suffice for our purposes -- to show that the induced
corrections on the $\gamma$ extraction are exceedingly small.  The
obtained result will also give us better understanding of the correct
EFT results with resummed $\log(m_b/M_W)$, which we will perform
next. The resummation is achieved by first integrating out $t$ and $W$
at $\mu\sim M_W$ and matching onto the effective theory with $b$, and
$c,s,d,u$ quarks. We will then evolve the Wilson coefficients down to
the scale $\mu\sim m_b$ using the renormalization-group (RG).

\subsection{The result without resummations}
We first evaluate the box diagram at $\mu\sim M_W$, treating $t$ and
$b$ quarks as massive and $u,c$ and $d,s$ quarks as massless, and set
all external momenta to zero (including the external $b$-quark
momentum). This will give us the local part of the EFT contributions
with unresummed Wilson coefficients. Because of the double GIM
mechanism, acting on both the internal up-quark and down-quark lines,
the leading contribution is proportional to $x_t y_b$, where
$x_t\equiv m_t^2/M_W^2, y_b\equiv m_b^2/M_W^2$. This is easy to see by
expanding the matrix element for the box-diagram correction to the
$B\to D K$ decay in terms of the quark masses,
\begin{equation}\label{Abox}
\begin{split}
A_{\rm box} = \sum_{u_i=u,c,t} \sum_{d_j=d,s,b} &\frac{G_F^2}{2}\lambda_{u_i}^{b\to s}
\lambda_{d_j}^{u\to c} \Big\{ A_1 M_W^2 +A_2 m_{d_j}^2 +A_3 m_{u_i}^2 +\\
&+ A_4 m_{u_i}^2 m_{d_j}^2 +\cdots \Big\} \times
\langle DK^- | (\bar s b)_{V-A}(\bar c  u)_{V-A} |B^-\rangle.
\end{split}
\end{equation}
The CKM factors $\lambda_{d_j}^{u\to c} = V_{u d_j}^* V_{c d_j}$ and
$\lambda_{u_i}^{b\to s} = V_{u_i b} V_{u_i s}^*$ are associated with
the flavor transitions on the internal down- and up-quark lines in
Fig. \ref{fig:ewcorr} (left), respectively.  The contributions in the
first line, proportional to $M_W^2$, $m_{d_j}^2$ and $m_{u_i}^2$,
vanish because either $\sum_{u_i=u,c,t} \lambda_{u_i}^{b\to s}=0$ or
$\sum_{d_j=d,s,b} \lambda_{d_j}^{u\to c}=0$.

Ignoring nonlocal contributions (see below), the box diagram with $b$
and $t$ quark massive and all the other quarks massless therefore
matches onto the effective Hamiltonian~\eqref{H1}.  This amounts to a
matching calculation where $t$ and $b$ quarks are integrated out
simultaneously at $\mu\sim M_W$ and results in a change $\Delta C_2$
of the Wilson coefficient $C_2$ in Eq.~\eqref{H1}, given by
\begin{equation}
\begin{split}\label{DeltaC2}
\Delta C_2& = \frac{\alpha}{4\pi\sin^2\theta_w}\frac{V_{tb}V_{ts}^*
  V_{ub}^*}{V_{us}^*} \hat C(x_t,y_b) = -
\left|\frac{\alpha}{4\pi\sin^2\theta_w}\frac{V_{tb}V_{ts}V_{ub}}{V_{us}}\right|
\hat C(x_t,y_b) e^{i\gamma}. 
\end{split}
\end{equation}
The Wilson coefficient $C_2$ in~\eqref{2} receives a similar
correction but with the same weak phase as the ${\mathcal O}(G_F)$
term. Thus the correction does not contribute to $\delta\gamma$ and we
neglect it, cf. Eqs. \eqref{btocsbaru}, \eqref{btousbarc}. The result
of our calculation agrees with the result extracted
from~\cite{Inami:1980fz} and reads
\begin{equation}\label{eq:Chatfull}
\hat C_\text{full}(x_t,y_b) = \frac{x_t \, y_b}{8} \bigg[
  \frac{9}{(x_t-1)(y_b-1)} + \bigg(
  \frac{(x_t-4)^2}{(x_t-1)^2(x_t-y_b)} \log x_t + (x_t \leftrightarrow
  y_b) \bigg) \bigg] \, .
\end{equation}
Note that the loop function $\hat C_\text{full}(x,y)$ vanishes if
either $x\to 0$ or $y\to 0$.  This proves that the only nonzero
contribution in \eqref{Abox} is $A_4\propto x_t y_b$. In fact, it is a
very good approximation to keep in this result only the $\log y_b$
enhanced contribution,
\begin{equation}\label{eq:Chat}
\hat C(x_t,y_b) = 2 y_b \log y_b+{\mathcal O}(y_b)\,,
\end{equation}
where the finite terms amount to an ${\mathcal O}(10\%)$
correction. Using the values for the CKM matrix elements from the
CKMfitter collaboration~\cite{Charles:2011va} and further input
from~\cite{Beringer:1900zz}, we find
\begin{equation}
\begin{split}\label{eq:unresum}
\Delta C_2&= (5.3\pm 0.3) \cdot 10^{-8} \times e^{i\gamma} \, ,
\end{split}
\end{equation}
where the error shown is only due to the CKM elements. 

The Wilson coefficient $\hat C(x_t, y_b)$ contains the unresummed
large logarithm $\log y_b$.  The logarithm is multiplied by $2y_b$ and
would vanish in the limit of zero $b$ quark masses. However, since the
Wilson coefficient $\hat C(x,y)$ starts only at ${\mathcal O}(y_b)$,
the term with $\log y_b$ represents a large correction. In the next
subsection we therefore perform a resummation of this logarithm.

\begin{figure}
    \centering
\raisebox{2.8cm}{$1)$~}\includegraphics[width=4cm]{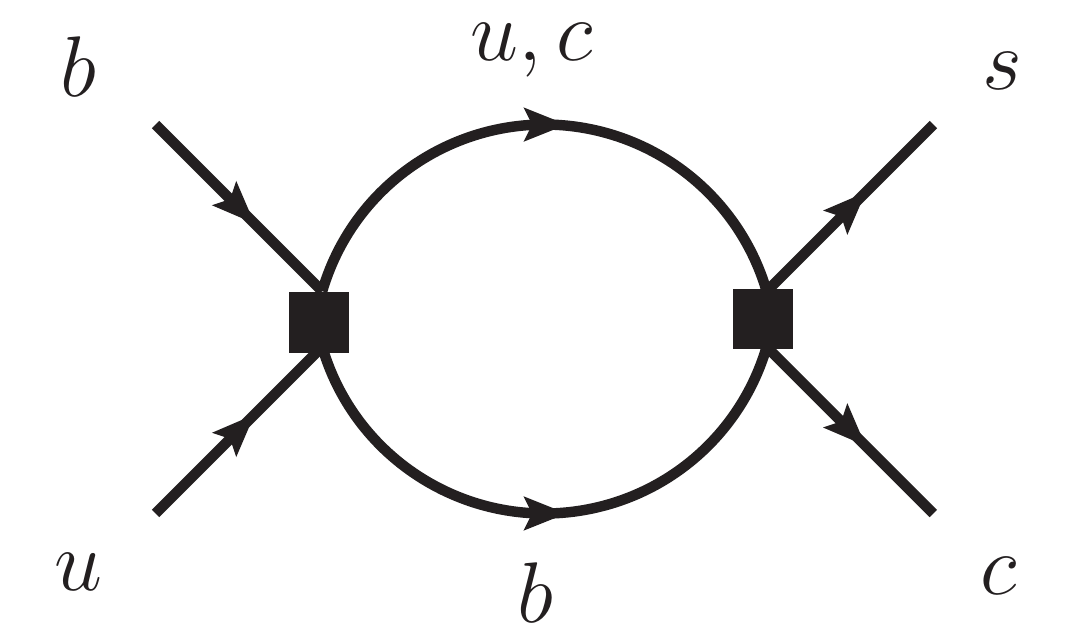} ~~~~~
\raisebox{2.8cm}{$2)$~}\includegraphics[width=4cm]{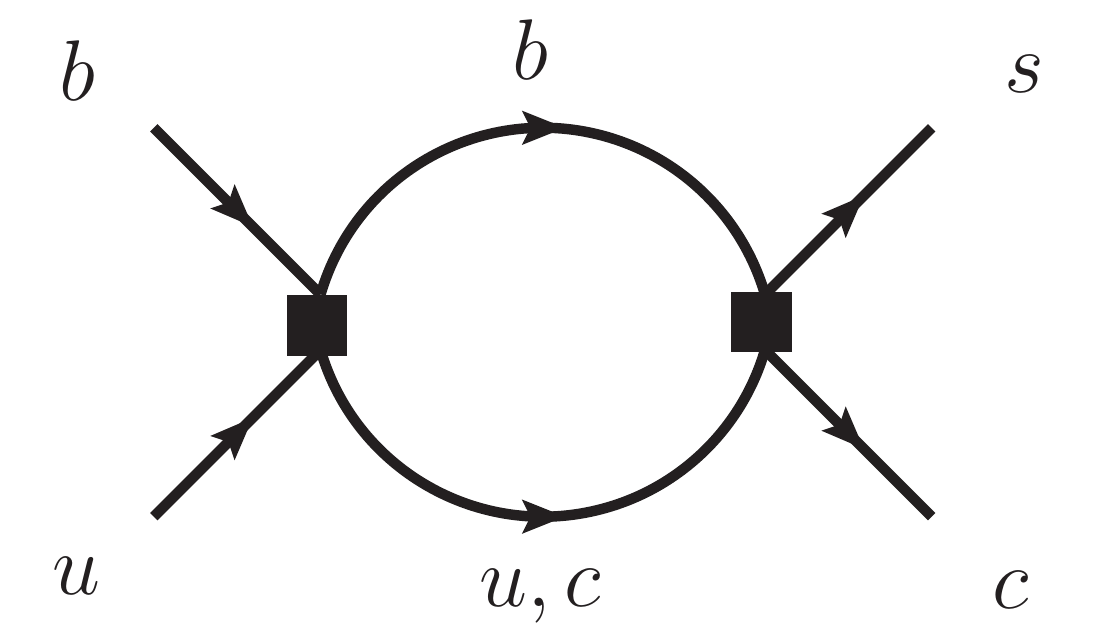} ~~~~~ 
\raisebox{2.8cm}{$3)$~}\includegraphics[width=4cm]{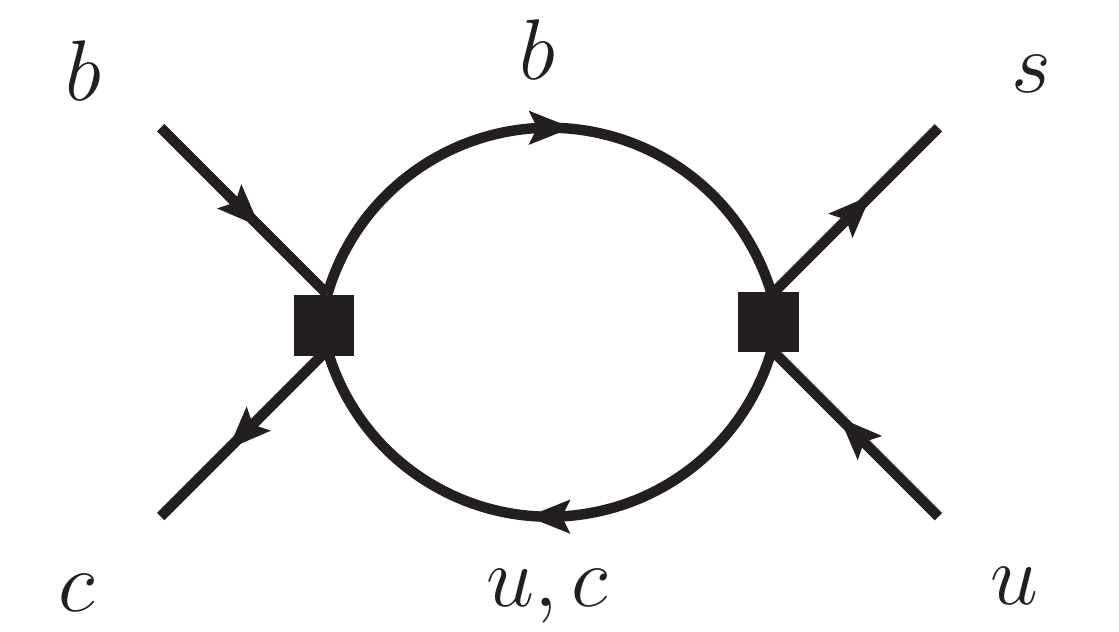} 
    \caption{ The double insertion $T\{Q_1,Q_1\}$, diagram 1), and
      $T\{Q_2,Q_2\}$, diagrams 2) and 3), contributing to the mixing
      into $\tilde Q_2$.  }
\label{fig:matching2}
 \end{figure}

\begin{figure}
    \centering
\raisebox{2.8cm}{$1)$~}\includegraphics[width=4cm]{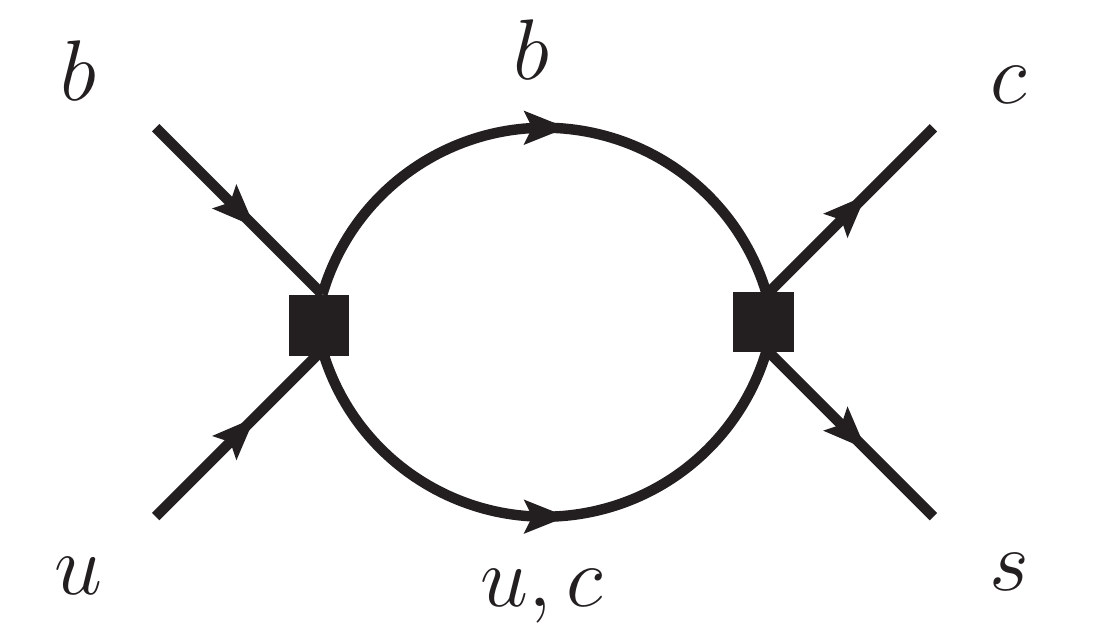} ~~~~~
\raisebox{2.8cm}{$2)$~}\includegraphics[width=4cm]{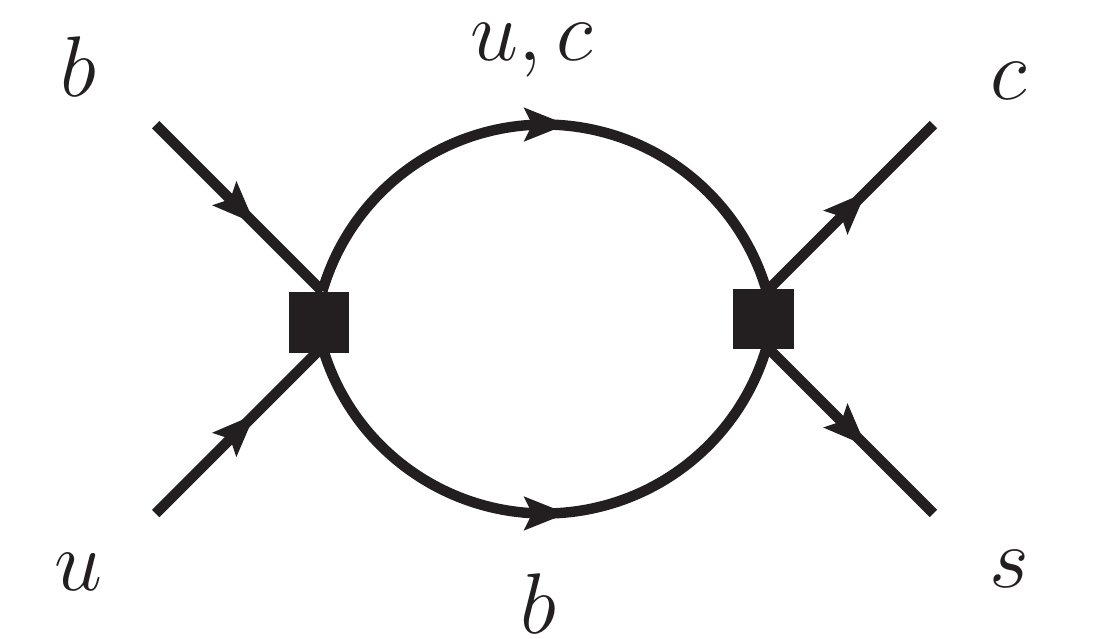} 
    \caption{ The double insertions $T\{Q_1,Q_2\}$ contributing to the
      mixing into the operator $\tilde Q_1$.}
\label{fig:matching3}
 \end{figure}

\subsection{The resummed result}

In order to resum $\log(m_b/M_W)$ we need to explicitly keep the
hierarchy of scales, $m_b\ll M_W$, in the construction of the
effective theories.  For $\mu>M_W$ one has the full SM, for
$m_b<\mu<M_W$ one has an effective theory with massless $b$ and
$c,s,d,u$ quarks but no top quark, while below $m_b$ there is an
effective theory with only the light quarks, $c,s,d,u$.

In the matching at $\mu\sim M_W$ the top quark and the $W,Z$ bosons
are integrated out, while the massless bottom quark is still a
dynamical degree of freedom also in the effective theory -- this is
the main difference to the previous subsection. Integrating out the
$W$ at tree level in electroweak counting generates the effective
Hamiltonians \eqref{1}, \eqref{2} and its variants with the
replacements $b,s\to d,s,b$ and $\bar cu \to \bar cc, \bar uu$. 

The contribution proportional to $y_b$ now vanishes at the electroweak
scale to the order considered. However, this contribution will be
generated by mixing of two insertions of dimension-six operators below
the electroweak scale. It is therefore useful to introduce the
following Hamiltonian describing the five-flavor effective theory
\begin{equation}\label{eq:H5}
\begin{split}
{\cal H}_{\rm eff}^{f=5} & = \frac{G_F}{\sqrt2} \sum_{\substack{u_{1,2}=u,c\\d_{1,2}=s,d,b}}
  V_{u_1d_2} V^*_{u_2d_1} \sum_{i,j = 1}^2 C_i(\mu) Z_{ij }Q_j^{(u_1
    d_2; d_1 u_2)} \\
& + 2G_F^2 V_{cb} V_{us}^* \cdot \bigg| \frac{V_{tb} V_{ts}
  V_{ub}}{V_{us}} \bigg| e^{i\gamma} \bigg[ \sum_{i,j,k=1}^2 C_i C_j
\hat Z_{ij,k} \tilde Q_k +\sum_{l,k=1}^2 \tilde C_l \tilde Z_{lk}
\tilde Q_k \bigg]  \, , 
\end{split}
\end{equation}
where we used $V_{tb} V_{ts}^* = - V_{cb} V_{cs}^* + {\cal
  O}(\lambda^2)$, with $\lambda = |V_{us}| \simeq 0.23$ (numerically,
this replacement is valid up to a three-permil correction). Moreover,
we denoted the usual four-quark operators by
\begin{equation}
  Q_1^{(u_1 d_2;d_1 u_2)} = (\bar u_1 d_2)_{V-A} (\bar d_1
  u_2)_{V-A} \, , \quad Q_2^{(u_1 u_2;d_1 d_2)} = (\bar
  u_1 u_2)_{V-A} (\bar d_1 d_2)_{V-A}\, ,   
\end{equation}
and defined
\begin{equation}
  \tilde Q_1 = \frac{m_b^2}{\mu^{2\epsilon} g_s^2} (\bar s
  u)_{V-A}(\bar c b)_{V-A} \, , \quad \tilde Q_2 =
  \frac{m_b^2}{\mu^{2\epsilon} g_s^2} (\bar s b)_{V-A}(\bar c u)_{V-A}
  \, . \label{tildeQ12}
\end{equation}
The last two operators denoted by a tilde are formally of dimension
eight because of the $m_b^2$ factor. They have the same four-quark
structure as the leading power operators $Q_{1,2}$ so that their
contributions could be absorbed by redefining the Wilson coefficients
$C_{1,2}$ allowing them to be complex. It is more practical, however,
to keep the Wilson coefficients real and split-off explicitly the
contributions to the effective Hamiltonian that carry the extra weak
phase as we did in \eqref{eq:H5}. Note that in the second line in
Eq.~\eqref{eq:H5} we neglect all the ${\mathcal O}(G_F^2)$ terms with
the same weak phase as the ${\mathcal O}(G_F)$ terms in
Eqs. \eqref{1}, \eqref{2}, since these are not relevant for
calculating $\delta \gamma$. We also neglect the six-quark operators
which arise from integrating out the $W$ boson and the top quark, as
they are suppressed by an additional factor of $1/M_W^2$. 

The dimension-eight Wilson coefficients at the electroweak scale
vanish to leading order. The mixing of double insertions of
dimension-six operators into $\tilde Q_{1,2}$ will generate
non-vanishing Wilson coefficients $\tilde C_{1,2}(\mu)$ below the
electroweak scale. The inverse powers of $g_s$ in the definition of
$\tilde Q_{1,2}$ in \eqref{tildeQ12} take into account that we will
sum the leading logarithms proportional to the strong coupling
constant.

Let us now look at some of the contributing terms in more detail. The sum
of the two diagrams denoted by 2) in Fig.~\ref{fig:matching2} yields
\begin{equation}
\begin{split}
  & C_2^2 V_{cb} V_{us}^* \big( |V_{ub}|^2 + V_{cs}^* V_{cb}
  \frac{V_{ub}^*}{V_{us}^*} \big) \langle Q_2 Q_2 \rangle_\text{div} = C_2^2 V_{cb} V_{us}^* \big(
  |V_{ub}|^2 + |V_{cb}|^2 \frac{V_{cs}^*V_{ub}^*}{V_{cb}^*V_{us}^*}
  \big) \langle Q_2 Q_2 \rangle_\text{div} \\
  &=  C_2^2 V_{cb} V_{us}^* \big(
  |V_{ub}|^2 + |V_{cb}|^2 \bigg| \frac{V_{cs}V_{ub}}{V_{cb}V_{us}}
  \bigg| e^{i\gamma} \big) \langle Q_2 Q_2 \rangle_\text{div} = C_2^2 V_{cb} V_{us}^* 
  \bigg| \frac{V_{cb} V_{cs} V_{ub}}{V_{us}} \bigg| e^{i\gamma} \langle Q_2 Q_2 \rangle_\text{div} +\cdots,\label{C2_diagr3}
\end{split}
\end{equation}
where $\langle Q_2 Q_2 \rangle_\text{div}$ is the common divergence of
the two diagrams, which is independent of the light-quark masses. In
the last step we kept only the term proportional to the factor with a weak phase,
which is the only contribution entering the shift $\delta\gamma$.  The
Lorentz and color structure of $\langle Q_2 Q_2 \rangle_\text{div}$ is
the same as of $\tilde Q_2$, so that this gives the anomalous
dimension of the double insertion mixing into $\tilde Q_2$. The sum of
the two diagrams denoted by 1) in Fig.~\ref{fig:matching2} is similar
to the first case, Eq. \eqref{C2_diagr3}, but with the replacement
$C_2\to C_1$, $Q_2\to Q_1$. The sum of the two diagrams denoted by 3)
in Fig.~\ref{fig:matching2} yields
\begin{equation}
\begin{split}
  & C_2^2 V_{cb} V_{us}^* \big( |V_{ub}|^2 + |V_{cb}|^2 \big) \langle Q_2 Q_2 \rangle_\text{div}, 
\end{split}
\end{equation}
and does not carry a weak phase.  As such it does not contribute to
$\delta \gamma$ and can be discarded.  There are also four additional
diagrams, shown in Fig.~\ref{fig:matching3}, which lead to the mixing
of double insertions into the Fierz-transformed operator $\tilde
Q_1$. 

To obtain the contributions of double ${\cal H}_{\rm eff}^{f=5}$
insertions to the running of $\tilde Q_{1,2}$ we thus only need to
compute the diagrams denoted by 1) and 2) in Fig.~\ref{fig:matching2},
with a double insertion of $Q_1$ and $Q_2$, respectively, plus two
additional diagrams with an insertion of $Q_1$ and then $Q_2$ at each
of the two weak vertices, cf. Fig.~\ref{fig:matching3}. We expand
$\hat\gamma_{i,j;k} = \frac{\alpha_s}{4\pi} \hat\gamma_{i,j;k}^{(0)} +
\ldots$, where $i,j$ denote the $Q_{1,2}$ insertions, and $k$ is the
labeling of the $\tilde Q_k$ operators. Extracting
$\hat\gamma_{i,j;k}^{(0)}$ from the one-loop divergence of the double
insertion (see, for instance,~\cite{Brod:2010mj} for details), our
calculation yields
\begin{equation}
\begin{split}
\hat\gamma_{1,1;2}^{(0)} = \hat\gamma_{2,2;2}^{(0)} = - 8 \, , \quad
\hat\gamma_{1,2;1}^{(0)} = \hat\gamma_{2,1;1}^{(0)} = 8 \, , 
\end{split}
\end{equation}
with all the remaining entries either vanishing or not contributing.
The initial conditions for the dimension-six Wilson coefficients are
given by $C_{1}(\mu_W) = 1$, $C_{2}(\mu_W) = 0$ to leading
order~\cite{Buchalla:1995vs}. Expanding $\tilde C_k = \tilde C_k^{(0)}
+ {\mathcal O (\alpha_s)}$, we find $\tilde C_k^{(0)}(\mu_W)=0$ at
leading order. A nonvanishing value will be induced by RG running for
$\mu < \mu_W$, which we compute by solving
\begin{equation}\label{eq:rgeinhom}
\mu \frac{d}{d\mu} \tilde C_k = \sum_l \tilde C_l \gamma_{lk} +
\sum_{ij} C_i C_j \hat\gamma_{ij,k} \, ,
\end{equation}
where $\gamma_{lk}$ is the well-known anomalous dimension for the
mixing of the $Q_{1,2}$ operators, 
\begin{equation}
\gamma_{lk} = 
\begin{pmatrix}
-2&6\\
6&-2
\end{pmatrix}\, .
\end{equation}

It is advantageous to go to the diagonal basis of the current-current
operators, by defining
\begin{equation}
\begin{split}
Q_\pm = \frac{1}{2} \big( Q_1 \pm Q_2\big) \, ,
\quad \tilde Q_\pm =
\frac{1}{2} \big( \tilde Q_1 \pm \tilde Q_2\big) \, .
\end{split}
\end{equation}
In this way Eq.~\eqref{eq:rgeinhom} gets rewritten as a homogeneous
equation~\cite{Herrlich:1996vf}, for which the standard techniques of
obtaining closed expressions for the RG evolution apply. The
transformed LO anomalous dimensions and the Wilson coefficients
are~\cite{Brod:2010mj}
\begin{equation}
\begin{split}
\gamma^{'(0)} = R \gamma^{(0)} R^{-1} \, , 
\qquad \hat
\gamma_{ij;k}^{'(0)} = R_{im} R_{jn} \hat \gamma_{mn;l}^{(0)} R_{lk}^{-1} ,
\qquad C^{'(0)} = \big( R^{-1} \big)^T C^{(0)}
\, ,
\end{split}
\end{equation}
where
\begin{equation}
R = \frac{1}{2}
\begin{pmatrix}
1&1\\
1&-1
\end{pmatrix}\, .
\end{equation}
By explicit calculation we find
\begin{equation}
\begin{split}
\hat\gamma_{+,+;-}^{'(0)} = 8 \, , \quad \hat\gamma_{-,-;+}^{'(0)} = - 8 \, , 
\end{split}
\end{equation}
while the remaining entries are zero. Defining $D_+ \equiv (C_-, \tilde
C_+ / C_-)^T$, the renormalization-group equations for $\tilde C_+$
and $C_-$ can be combined into
\begin{equation}
\begin{split}
\mu \frac{dD_+}{d\mu} = \gamma_{D_+} \cdot D_+ \, ,
\qquad {\rm where~~~}
\gamma_{D_+} = 
\begin{pmatrix}
\gamma_-&0\\
\hat\gamma_{-,-;+}&\gamma_+ - \gamma_- + 2 \gamma_m - 2 \beta
\end{pmatrix}\, .
\end{split}
\end{equation}
We obtain the corresponding solution for $\tilde C_-$ and $C_+$ by
exchanging the subscripts $+ \leftrightarrow -$. Note that we have
also included the running of the mass and the coupling constant
related to the factor $m_b^2/g_s^2$ in the definition of the operators
$\tilde Q_k$, given by the anomalous dimension of the quark mass
$\gamma_m$ and the QCD beta function $\beta$. Transforming back to the
original basis, we find numerically
\begin{equation}
\{\tilde C_1(m_b), \tilde C_2(m_b)\} =
\{0.03, 0.31\}\, ,
\end{equation}
where we used $\alpha_s(M_Z) = 0.1184$~\cite{Beringer:1900zz} and
$m_b(m_b) = 4.163\,\text{GeV}$~\cite{Chetyrkin:2009fv}. Note that the
RG running has now also induced a nonzero correction to $C_1$ in
\eqref{H1}, in contrast to the unresummed result. We used the
mathematica package ``RunDec''~\cite{Chetyrkin:2000yt} for the
numerical running of the strong coupling constant.

Finally, at the bottom-quark scale we need to calculate the $B\to DK$
matrix elements using our EFT Hamiltonian \eqref{eq:H5} in order to
obtain the shift $\delta\gamma$.
This will give the leading $y_b$ behavior with resummed logarithms.
We write the matrix elements suggestively as
\begin{equation}
\begin{split}
\sum_{k}
\Delta C_k(\mu_b) \langle Q_k \rangle (\mu_b) =
2\sqrt{2}G_F \left |
\frac{V_{tb}V_{ts}V_{ub}}{V_{us}} \right | e^{i\gamma}\left[  \sum_{i,j = 1}^{2} C_i(\mu_b) C_j(\mu_b)  \langle Q_i Q_j
  \rangle(\mu_b)\right.\\
   \left. + \sum_{i=1,2} \tilde C_i(\mu_b) \langle \tilde Q_i
  \rangle (\mu_b) \right]\, .
\end{split}
\end{equation}
Here we expand $\Delta C_k = \frac{4\pi}{\alpha_s} \Delta C_k^{(0)} +
{\mathcal O}(1)$; note that in this way the artificially inserted
factor of $1/g_s^2$ in the definition of $\tilde Q_k$~\eqref{tildeQ12}
is canceled. At LO it is not necessary to compute the double
insertions $\langle Q_i Q_j \rangle$ since these are loop suppressed, and therefore we effectively obtain the matching condition 
for the Wilson coefficients of the local operators \eqref{H1}
\begin{equation}
 \Delta C_k^{(0)}(\mu_b) = 2 m_b^2  \frac{\sqrt{2}G_F}{16\pi^2} \left |
\frac{V_{tb}V_{ts}V_{ub}}{V_{us}} \right | e^{i\gamma}  \tilde C_k^{(0)}(\mu_b) \, .
\end{equation}
Numerically, we find
\begin{equation}\label{DeltaC12-resum}
|\Delta C_1| = (4.5 \pm 0.2) \cdot 10^{-9} \, , \quad |\Delta C_2| =
(4.3 \pm 0.2) \cdot 10^{-8} \, ;
\end{equation}
the errors reflect the uncertainty in the electroweak input
parameters. This should be compared to the unresummed result
Eq.~\eqref{eq:unresum}. Expanding the solution of the
renormalization-group equations around $\mu=M_W$ and expressing $G_F$
in terms of the weak mixing angle we recover exactly the logarithm in
Eq.~\eqref{eq:Chat}:
\begin{equation}\label{DeltaC12-expanded}
\Delta C_1 = 0 \, , \quad \Delta C_2 = 2 y_b 
\frac{\alpha}{16 \pi \sin^2\theta_w} ( -4 \log y_b) \, .
\end{equation}

\section{Conclusions}
\label{Conclusions}
The determination of the SM weak phase $\gamma$ from the $B\to DK$
decays has a very small irreducible theoretical error which is due to
one-loop electroweak corrections. In this paper we have estimated the
resulting shift in $\gamma$. Treating $m_b\sim M_W$ or resumming logs
of $m_b/M_W$ gives in both cases an estimated shift $\delta \gamma
\sim 2\cdot 10^{-8}$, keeping only the local operator contributions at
the scale $\mu \sim m_b$. It is unlikely that the neglected non-local
contributions, which come with the same CKM suppression as the local
contributions, would differ from the above estimate by more than a
factor of a few. We can thus safely conclude that the irreducible
theoretical error on the extraction of $\gamma$ from $B\to DK$ is
$|\delta \gamma|\lesssim {\mathcal O}(10^{-7})$.

{\bf Acknowledgements:} We would like to thank Dan Pirjol for
collaboration at earlier stages of this work and many discussions and
helpful comments on the manuscript. J.B. and J.Z. were supported in
part by the U.S. National Science Foundation under CAREER Grant
PHY-1151392. J.B. would like to thank the Kavli Institute for
Theoretical Physics for hospitality during a visit when part of this
work was completed. This research was supported in part by the
National Science Foundation under Grant No. NSF PHY11-25915.

\end{document}